\newcommand{\bp}{\textbf{p}}
\newcommand{\bk}{\textbf{k}}
\newcommand{\bq}{\textbf{q}}

\documentclass[aps,prb,superscriptaddress,twocolumn,showpacs,amsmath,amssymb]{revtex4}
\usepackage{graphicx}% Include figure files
\usepackage{dcolumn}% Align table columns on decimal point
\usepackage{bm}% bold math
\usepackage{color}
\usepackage{here}
\definecolor{DarkBlue}{rgb}{0.1,0.1,0.5}
\definecolor{Red}{rgb}{0.9,0.0,0.1}
\definecolor{Green}{rgb}{0.0,0.99,0.0}

\begin{document}
\title{Aspects of Electron-Phonon Self-Energy Revealed from Angle-Resolved Photoemission Spectroscopy}
\author{W.S. Lee}
\affiliation {Department of Physics, Applied Physics, and Stanford
Synchrotron Radiation Laboratory, Stanford University, Stanford,
CA 94305}

\author{S. Johnston}
\affiliation {Department of Physics, University of Waterloo,Waterloo, Ontario, Canada N2L 3G1}

\author{T.P. Devereaux}
\affiliation {Department of Physics, University of Waterloo,Waterloo, Ontario, Canada N2L 3G1}

\author{Z.-X. Shen}
\affiliation {Department of Physics, Applied Physics, and Stanford Synchrotron
  Radiation Laboratory, Stanford University, Stanford, CA 94305}

\date{\today}% It is always \today, today,
             %  but any date may be explicitly specified

\begin{abstract}
Lattice contribution to the electronic self-energy in complex correlated oxides is a fascinating subject that has lately
stimulated lively discussions. Expectations of electron-phonon self-energy effects for simpler materials, such as Pd and Al, have
resulted in several misconceptions in strongly correlated oxides.
Here we analyze a number of arguments claiming that phonons cannot
be the origin of certain self-energy effects seen in high-$T_c$
cuprate superconductors via angle resolved photoemission
experiments (ARPES), including the temperature dependence, doping
dependence of the renormalization effects, the inter-band
scattering in the bilayer systems, and impurity substitution. We
show that in light of experimental evidences and detailed
simulations, these arguments are not well founded.

\end{abstract}

\pacs{Valid PACS appear here}% PACS, the Physics and Astronomy
                             % Classification Scheme.
%\keywords{Suggested keywords}%Use showkeys class option if keyword
                              %display desired
\maketitle

\section{Introduction}
The microscopic pairing mechanism of the high-$T_c$
superconductivity remains an unsolved question even after twenty
years of its discovery. Observations of a kink at around 40-70 meV
in the band dispersion along the diagonal of the Brillouin zone
(nodal diection) and a peak-dip-hump (PDH) structure at the zone
boundary by angle-resolved photoemission spectroscopy (ARPES)
\cite{Mode:Pasha,Lanzara:phonon,Mode:Kaminski,mode:Kim,mode:sato:BiFamily,
Tanja:B1g:PRL,ZJZhou:multiple_mode,non:doping_dependence_monolayer,
Mode:Norman_antinodal,Mode:Gromko,A.A.Kordyuk:kink_with_self_consistent,Borisenko:interband:Bi2212,Borisenko:interband:YBCO}
have drawn a great deal of recent attention as they may shed some
light on this problem. Although an agreement has been established
that the kink and PDH structure are signatures of the electrons
coupled to a sharp bosonic mode, it is still strongly debated
about the origin of this bosonic mode. Influenced by the fact that
the high-$T_c$ cuprate is a doped antiferromagnetic insulator, a
common belief is that this bosonic mode has a magnetic origin
\cite{A.A.Kordyuk:kink_with_self_consistent,Borisenko:interband:Bi2212,Borisenko:interband:YBCO,Mode:Gromko,
Mode:Kaminski,mode:Kim,Mode:Norman_antinodal,mode:sato:BiFamily}.
However, an alternative view is that the electron-phonon coupling
in such a doped-insulator can be very strong and anomalous because
of a number of unusual effects, such as poor screening, complex
structure as well as the interplay with correlations. Indeed,
oxygen related optical phonons have been invoked to explain the
temperature and doping dependence of the renormalization effects
\cite{Lanzara:phonon,Tanja:B1g:PRL,Tom:B1g:PRL,ZJZhou:multiple_mode,non:doping_dependence_monolayer}.
This idea of phonons being mainly responsible for this low energy
band renormalization effect observed by ARPES has stimulated
intense debate. There is currently no consensus whether a phonon,
a set of phonons, or a magnetic mode is the primary cause of the
band renormalization.

As mentioned, some important reasons to invoke phonon
interpretation of the ARPES data are: the presence of an universal
energy scale in all materials at all doping
\cite{Lanzara:phonon,Universal_velocity:Xingjian}, particularly in
the normal state of very low $T_c$ materials
\cite{Lanzara:Noraml_state_PDH}; the strong inferred momentum
dependence \cite{Tanja:B1g:PRL,Tom:B1g:PRL}; the existence of
multiple bosonic mode couplings \cite{ZJZhou:multiple_mode} and
the decrease in the overall coupling strength with increased
doping, interpreted as a screening effect, especially for phonons
with eigenvectors along the
c-axis\cite{non:doping_dependence_monolayer}. Yet, there is still
a widespread belief that phonons are not responsible for the kink
features. In the following sections, with a comprehensive look at
all experimental data as well as some recent simulations, we
address some of the criticisms that have been commonly used to
argue against the phonon interpretation. These include the
temperature and doping dependence of the renormalization effects,
inter-band scattering for bilayer system, and the ARPES
experiments on impurity substituted Bi2212 crystal,
Bi$_2$Sr$_2$Ca(Cu$_{2-x}$M$_x)$O$_{8+\delta}$ with M = Zn or Ni.
Our goal is to clarify these misconceptions as being due to
oversimplifying the effects of electron-phonon coupling in
cuprates as well as other strongly correlated transition metal
oxides.
\section{Aspects of The Electron-Phonon Self-energy}
\subsection{Temperature Dependence}
In the standard treatment of electron-phonon coupling effects, the
Debye temperature sets a characteristic temperature scale, which
is well above $T_c$ in conventional superconducting materials.
However in the cuprates and other low Fermi energy systems, these
two energy scales can be comparable. As a result, the temperature
dependence of phonon induced self-energies can be very different
from that of conventional superconductors. According to the ARPES
measurements on Bi2212 system, the band renormalization in the
antinodal region (peak-dip-hump structure) shows a dramatic
superconductivity-induced enhancement when the system goes through
a phase transition from the normal state to the superconducting
state. It has been argued that only a mode which emerges in the
superconducting state and vanishes in the normal state can explain
this temperature-dependent renormalization effect
\cite{Mode:Kaminski,mode:Kim,mode:sato:BiFamily,Mode:Norman_antinodal}.
Phonons are thereby excluded.

The sharpness of the renormalization effects due to
electron-phonon coupling is strongly temperature dependent, given
by the fact that $T_c$ of optimally-doped Bi2212 is close to 100K.
To demonstrate this temperature dependence, we consider the normal
state (120 K) and superconducting state (10 K) of a d-wave
superconductor coupled to a 36 meV $B_{1g}$, 55meV oxygen
$A_{1g}$, and 70 meV breathing phonons
\cite{Tom:B1g:PRL,Screening:Tom}. The electron-phonon coupling for
the $B_{1g}$ and breathing phonons are those used in Ref.
\raisebox{-1.2ex}{\Large\cite{Tom:B1g:PRL}}. The $A_{1g}$ modes involve c-axis
motions of both planar and apical oxygens, and have two branches around 55 and
80 meV. The apical electron-phonon coupling, derived in Ref. \raisebox{-1.2ex}{\Large
\cite{Screening:Tom}}, involves a charge-transfer from the apex
oxygen into the CuO$_{2}$ plane via the Cu 4s orbital, the same
pathway that gives rise to bi-layer splitting. However, for
simplicity, the apical electron-phonon coupling is treated as a
constant in the calculations presented in this paper. The reason
to include three modes in the calculation was inspired by the
earlier success of the two-mode calculation\cite{Tanja:B1g:PRL} as
well as the recent discovery of multiple mode
coupling\cite{ZJZhou:multiple_mode,non:doping_dependence_monolayer}.
For this calculation, the tight-binding band structure described
in Ref. \raisebox{-1.2ex}{\Large\cite{tight_binding:Norman}} has
been used. The real part and imaginary part of the self-energy
$\Sigma({\bf k},\omega)$ and the spectral functions $A({\bf
k},\omega)$ at $k=(0,\pi)$ are then obtained within weak coupling
Eliashberg formalism \cite{Tom:B1g:PRL} and plotted in Fig.
\ref{Fig:T_dep}. Details of the calculations are presented in the Appendix.

At high temperature, both Re$\Sigma({\bf k},\omega)$ and
Im$\Sigma({\bf k},\omega)$ do not exhibit a sharp renormalization
feature as shown by the dashed curves in Fig. \ref{Fig:T_dep} (a)
and Fig. \ref{Fig:T_dep}(b), respectively. This demonstrates that
the thermal broadening effect smears out the sharp renormalization
features; in addition, broadening effects due to additional many
body effects would smear out the renormalization features further.
Thus, one should not expect to observe any sharp renormalization
features at $k=(0,\pi)$ in the normal state ($\sim$ 100K) from the
electron-phonon coupling. In the superconducting state, the
renormalization features of the self-energy become much sharper,
due to smaller thermal broadening effect as well as the opening of
a superconducting gap. Consistent with the optimally-doped Bi2212
and Bi2223 measurements
\cite{Mode:Kaminski,mode:sato:BiFamily,Tanja:B1g:PRL,weishen:Bi2223},
the PDH structure of the spectral function at $k=(0,\pi)$ emerges
at low temperature and disappears at high temperature (normal
state), as illustrated in Fig. \ref{Fig:T_dep}(c) and Fig.
\ref{Fig:T_dep}(d), respectively. While this behavior is generally
expected for any phonon, we note that in addition, the self-energy
from electron-phonon couplings which involve momentum transfers
within and between anti-nodal regions of the Fermi surface, such
as the apex $A_{1g}$ and $B_{1g}$ phonons, are greatly enhanced
for all k-points due to the large density of state enhancements in
these regions via the opening of a $d$-wave gap. A detailed
momentum dependence of the renormalization effects in the normal
state and superconducting state due to the coupling to the
B$_{1g}$ phonon has been discussed in Ref.
\raisebox{-1.2ex}{\Large\cite{Tanja:B1g:PRL}} and Ref.
\raisebox{-1.2ex}{\Large\cite{Tom:B1g:PRL}}. Furthermore, both the
dispersion kink and the PDH structure in the nodal region have
been clearly observed in the normal state when measured at a low
temperature on samples with a lower $T_c$
\cite{Lanzara:Noraml_state_PDH}. This lends further support to the
strongly temperature dependent renormalization features due to
electron-phonon coupling.

\begin{figure} [t]
\includegraphics [width=3.25 in, clip]{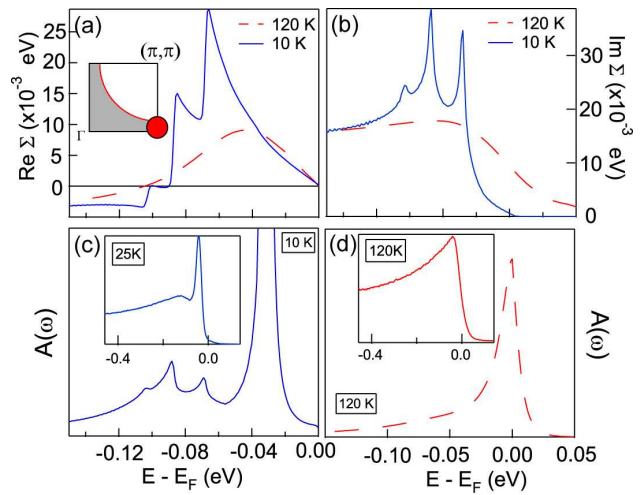}
\caption{\label{Fig:T_dep} The calculated (a) real part,
Re$\Sigma$, (b) imaginary part of the self-energy, Im$\Sigma$ ,
and the corresponding spectral functions, A(k,$\omega$) in (c)
normal state and (d) superconducting state. An extra 5 meV is
added to the imaginary part of the self energy in 120K simulation
to account for the thermal broadening of the quasi-particle life
time. The location for this calculation is indicated in inset of
(a) by a red dot with a red curve representing the FS. Insets of
(c) and (d) are the data of optimally-doped Bi2223 system
($T_c$=110K) taken at the normal state (120K) and superconducting
state(25K)\cite{weishen:Bi2223}, respectively.}
\end{figure}

\subsection{Doping Dependence}
Another problematic statement against the phonon scenario stems
from the apparent strong doping dependence of the kink position
and strength. Based on the wisdom for conventional good metals,
phonons should not have a strong doping dependence, either in
frequency of the mode or in strength of the coupling. This is not
necessary valid for layered, doped insulators with strong
correlation effects, such as cuprates where doping dramatically
changes the metallicity and the ability of electrons to screen
charge fluctuations. We first note that many experiments on
various cuprates have reported strongly doping dependent anomalies
for several phonons, which implies a strongly doping dependent e-p
coupling for these phonons. For example, from inelastic neutron
scattering measurements, the breathing mode, half-breathing mode,
and the bond-stretching modes exhibit prominent doping dependence
of dispersion and energy renormalizations \cite{L. Pintschovius:neutron:el_ph,Reznik:Pinstchovius:neutron:el_ph_nature}.
In Raman and infrared spectroscopy, the Fano lineshapes of phonon
modes in $B_{1g}$ and $B_{1u}$ symmetry show strong doping
dependences \cite{M. Opel:Ramman:doping_dep_phonon}. Furthermore,
the strength of the phonon energy shift and linewidth variation
across $T_c$ also changes strongly with doping
\cite{Hardy:YBCO:B1g:Raman}.

\begin{figure}[t]
\includegraphics [ width=3.25 in, clip]{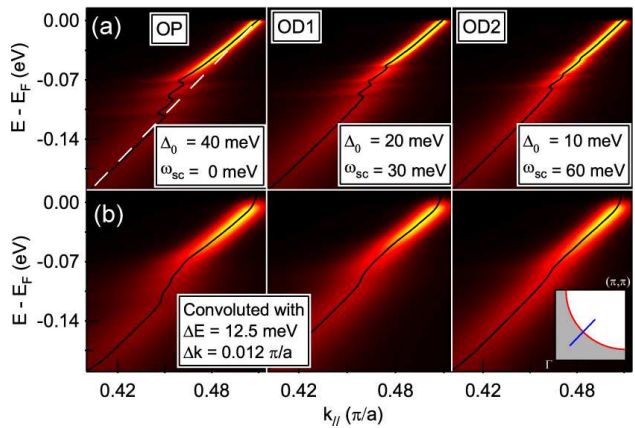}
\caption{\label{Fig:doping_dep} The intensity plots of the (a)
spectral functions without resolution convolution and (b)
resolution convoluted spectral functions in the superconducting
state (10K) along the nodal direction, as indicated in the inset
of (b) by the blue line. Black curves are the band dispersion
extracted from the maximum positions of the momentum distribution
curves, which cut the spectral functions at a fixed energy. The
MDC-derived dispersions in (a) exhibit three sharp "subkinks" due
to the coupling to the three phonon modes used in the model; while
in (b) the subkinks are washed out by the finite instrument
resolution effect leaving an apparent single kink in the band
dispersion. The white dashed line illustrates the bare band for
extracting Re$\Sigma$ shown in Fig. \ref{Fig:doping_dep_summary}
(a).}
\end{figure}

Second, the doping dependence of the renormalization effects to
the electronic self-energy is sophisticated as inferred by two
recent ARPES studies. One is the observation of multiple bosonic
mode couplings along the nodal direction
\cite{ZJZhou:multiple_mode}. The other is the doping dependence of
the c-axis screening effect to the coupling between the electron
and $c-$axis phonons. As proposed by Meevasana \emph{et
al.}\cite{non:doping_dependence_monolayer,Jan} and Devereaux
\emph{et al.}\cite{Screening:Tom}, for electron-phonon coupling at
long wavelengths, the screening becomes more effective at reducing
the coupling strength when the c-axis conductivity becomes more
metallic. Given these two results plus the variation of the
superconducting gap magnitude with doping, the doping dependence
of the kink energy is highly convoluted in Bi2212 whose
superconducting gap has an energy comparable to some of the
phonons.

In Fig. \ref{Fig:doping_dep}, we present the intensity plot of
calculated spectral functions demonstrating a doping dependence of
the dispersion kink in the superconducting state. The
superconducting gap size was set to be 40, 20, and 10 meV for the
optimally-doped and more overdoped systems. In addition, the
coupling strength of the breathing mode, whose appreciable
coupling is only for short wavelengths and large momentum
transfers\cite{L. Pintschovius:neutron:el_ph,Reznik:Pinstchovius:neutron:el_ph_nature},
remains unchanged for our doping dependence simulation; while, a
filter function, $\omega^2/(\omega^2+\omega_{sc}^2)$ with
different value of c-axis screening frequency $\omega_{sc}$ is
applied to the c-axis phonons (B$_{1g}$ and A$_{1g}$), to
simulate the doping-dependent coupling strength due to the change
of the c-axis screening effect
\cite{non:doping_dependence_monolayer,Jan}.
We note that although
this is a simplification, it represents the general behavior of
screening considerations for phonons involving small in-plane
momentum transfers. Full consideration of screening has been given
in Ref. \raisebox{-1.2ex}{\Large\cite{Screening:Tom}} and Ref.
\raisebox{-1.2ex}{\Large \cite{Jan}}. In addition, a component
$0.005 + \omega^2$ eV is added in the imaginary part of the
self-energy to mimic the quasiparticle life time broadening due to
electron-electron interaction.

\begin{figure}[t]
\includegraphics [ width=3.25 in, clip]{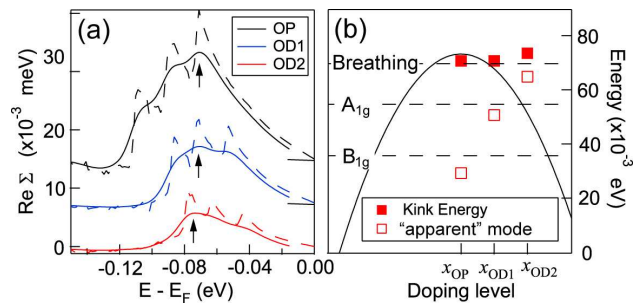}
\caption{\label{Fig:doping_dep_summary} (a) The Re$\Sigma$
extracted from Fig. 2(a) (dashed lines) and Fig. 2(b) (solid
lines) by subtracting a linear bare band (dashed line in Fig.
\ref{Fig:doping_dep}(a)) from the band dispersion. The arrows
indicate the maximum positions of the Re$\Sigma$ where the
``single" apparent kink in the band dispersion is usually defined.
(b) Summary of the doping dependence of the apparent kink energy
and the apparent mode energy extracted by assuming a single mode
scenario.}
\end{figure}

As shown in Fig. \ref{Fig:doping_dep}(a), the coupling to multiple
phonon modes induces several ``subkinks" in the dispersion. The
positions of these subkinks mostly correspond to the energies of
phonons plus the maximum d-wave SC gap, $\Delta_0$, even through
there is no gap along the nodal direction. This is because when
calculating the self-energy, one needs to integrate the
contributions from the entire zone, which makes the electrons in
the nodal region "feel" the presence of the gap. Furthermore,
revealed by the extracted real-part of the self-energy, Re$\Sigma$
(dashed curves in Fig. \ref{Fig:doping_dep_summary} (a)), the
dominant feature in Re$\Sigma$ for the OP case is induced by 36
meV B$_{1g}$ mode, while for the OD1 and OD2 case, the features of
the 55 meV $A_{1g}$ mode and 70 meV breathing mode gradually
out-weight the contribution from the B$_{1g}$ mode. This
demonstrates that the change of the SC gap magnitude and the
increasing screening effect to these phonons because of increased
doping alters the relative strength of the subkinks caused by
different modes.

To simulate the experimental data, we convoluted the spectral
functions shown in Fig. \ref{Fig:doping_dep}(a) with a typical
ARPES instrumental resolution: 12.5 meV in energy resolution and
0.012 $\pi/a$ in momentum resolution. As illustrated in Fig.
\ref{Fig:doping_dep}(b) and the extracted Re$\Sigma$ (solid curves
in Fig. \ref{Fig:doping_dep_summary}(a)), the subkinks are less
obvious and become a broadened ``single" kink in the dispersion
which is located at roughly the energy of the dominant phonon
feature determined by the maximum position of the Re$\Sigma$ (the
arrows in Fig. \ref{Fig:doping_dep_summary}(a)). The doping
dependence of the kink position is summarized as the solid symbols
in Fig. \ref{Fig:doping_dep_summary}(b). Assuming a single mode
scenario, one can obtain the ``doping dependence" of the mode
energy by subtracting out the SC gap size. However, we note that
this extracted ``apparent" mode energy does not match any of the
modes used in the model; instead, it is a average between the
dominant features (open symbols in Fig.
\ref{Fig:doping_dep_summary}(b). Clearly, since the kink energy is
a sum of the superconducting gap and a spectrum of bosonic modes,
it should not be taken as a precise measurement of the energy of
any particular bosonic mode. This casts doubts to the analysis of
the doping dependent properties of the kink in the nodal band
dispersion based on the single bosonic mode coupling scenario
\cite{A.A.Kordyuk:kink_with_self_consistent,Borisenko:interband:YBCO}.
More importantly, this illustrates the complex nature of lattice
effects in these oxides.

\subsection{Interband Scattering}
Borisenko \emph{et al.} \cite{Borisenko:interband:Bi2212,
Borisenko:interband:YBCO} observed that the scattering rate of the
bonding and antibonding bands along the nodal direction cross each
other near the energy of the Van Hove singularity, suggesting a
strong inter-band scattering between the bonding and antibonding
bands. They argued that only a mode with "odd" symmetry, such as
spin resonance mode, can mediate such inter-band scattering. The
question whether phonons can induce such inter-band scattering has
also been raised by these authors.

First, we note that recent high energy and momentum resolution
ARPES experiments on Bi2212 using low energy photons( $<$10 eV)
have better resolved the bilayer splitting at the nodal point
\cite{low_photon_energy:Ino}. However, as shown in Fig. 2 of Ref.
\raisebox{-1.2ex}{\Large\cite{low_photon_energy:Ino}}, the
scattering rate of the bonding and anti-bonding band does
\emph{not} exhibit a crossover behavior as reported by Borisenko
\emph{at al.}. The inconsistency of the data between the two
groups implies that more experiments and better analysis are
needed to verify whether this inter-band scattering effect is
genuine.

\begin{figure} [t]
\includegraphics [width=2.75 in, clip]{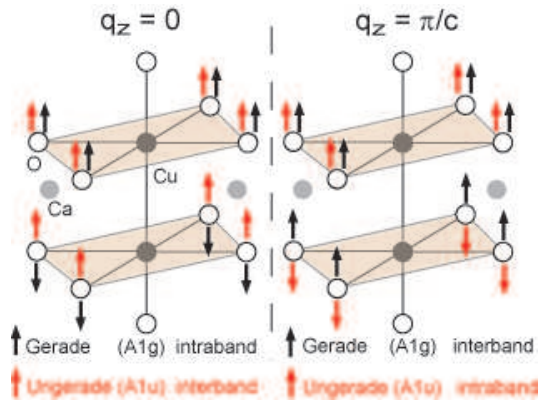}
\caption{\label{Fig:interband} The illustration of the gerade and
ungerade $c-$axis phonons. The eigenmode of the gerade (ungerade)
phonons is even (odd) with respect to the mirror plane between two
CuO$_2$ layers at $q_z=0$, while their definition swapped at
$q_z=\pi/c$. The black, grey, and white circles represent the Cu,
Ca, and O atoms, respectively.}
\end{figure}

Second, empirically, it has been known for over 15 years that
interband electron-phonon coupling in the cuprates is very large.
The evidence comes from the strong resonance profiles of many
Raman active phonons, which display large intensity
variations\cite{Heyen:interband_scattering_by_phonon}. This is
generally understood as a result of strong interband coupling,
whereby phonons can be brought in and out of resonances via tuning
of the incident photon energy \cite{Sherman}. Since, in general,
phonons can also provide momentum to scatter electrons along the
c-axis, direct inter bi-layer scattering can occur which involves
mixing of different symmetries of phonons. This can be viewed in a
simplified way even if we first neglect direct interband
scattering and consider a bilayer system coupled to $c-$axis
phonons. For $q_z=0$, a simple classification of c-axis modes is
possible:
\begin{widetext}
\begin{eqnarray}
H =&& \sum_{k, \sigma, \alpha=1,2}
\epsilon_\alpha(k)c_{k,\alpha,\sigma}^\dagger
c_{k,\alpha,\sigma}^{} +\frac{1}{2}\sum_{k,\sigma}t_{\perp}(k)
\left[ c_{k,1,\sigma}^\dagger c_{k,2,\sigma}^{} +
h.c.\right]\nonumber\\
&&+\sum_{k,q,\sigma,\alpha=1,2,\nu}g_{\nu,\alpha}(k,q)\left\{c_{k+q,\alpha,\sigma}^\dagger
c_{k,\alpha,\sigma}^{}\left[a_{\nu}(-q)+a_{\nu}^\dagger(q)\right]+h.c.\right\}\label{eq:Hamiltonian},
\end{eqnarray}
\end{widetext}
where $\alpha$ is the index for the electronic states of different
layers, $\epsilon_1(k)=\epsilon_2(k)$, $t_{\perp}(k)$ describes
the hopping of electrons between two layers, and the index $\nu$
can be either gerade or ungerade active c-axis modes, with
symmetry classification with respect to the displacement
eigenvectors to the inversion center of the cell, depicted in Fig.
\ref{Fig:interband}. After diagonalizing the first two terms by
canonical transformation, the electron-phonon coupling can be
recast as
\begin{eqnarray}
&&H_{e-ph}^{(g,u)}=\sum_{k,q,\sigma}g_{(g,u)}(k,q)\left\{\left[a_{(g,u)}(q)+a_{(g,u)}^\dagger(-q)\right]\right.\nonumber\\
&&\left.\times\left[c_{k+q,+,\sigma}^\dagger c_{k,(+,-),\sigma}^{}
+ c_{k+q,-,\sigma}^\dagger c_{k,(-,+),\sigma}^{} \right] +
h.c.\right\}.\label{eq:after_diagnol}.
\end{eqnarray}

We have used the $c_+$ and $c_-$ for the even and odd linear
combination of $c_1$ and $c_2$, and subscript $g$ and $u$ for the
gerade and ungerade mode, respectively. Thus for $q_{z}=0$, where
this classification is possible, gerade phonons induce intra-band
scattering (even channel), while the ungerade phonons mediate the
inter-band scattering process (odd channel) even without direct
electron-phonon coupling across the layers. Yet for $q_{z}=\pi/c$,
the classification inverts, where gerade modes become ungerade and
vice-versa, as illustrated in Fig \ref{Fig:interband}. Thus, even
in this simple case, modes at different $q_{z}$ contribute both to
intra and interband scattering, and the net weight of the coupling
appearing in the self energy is then largely determined by the
specific momentum structure $g(k,q)$. Since the self-energy
generally involves sums over $q_z$, and coupling directly of
electrons in adjacent layers via phonons are non-negligible,
clearly the inter-band scattering phenomena can not be used to
argue against the phonons being important to the electronic
states.

We also add a remark concerning the electron-phonon coupling
derived from Raman measurements \cite{footnote:B1g coupling
strength} in YBa$_{2}$Cu$_{3}$O$_{7}$ and Bi-2212 compared to that
obtained from ARPES. While one might naively expect the couplings
to be comparable from Raman and ARPES, we remark that this
situation is remarkably different if the coupling is strongly
moment dependent and whenever correlations are appreciable. Since
Raman measures phonons with net zero momentum transfer and ARPES
involves a sum over all transfers, a sizeable coupling difference
may be discernable. This is specifically the case for the $B_{1g}$
phonon, where scattering involving momentum transfers across the
necks of the Fermi surface near $(\pi,0)$\cite{Tom:B1g:PRL},
further enhanced via correlations\cite{Fu:B1g_and
correlation_enhance}, yields a strong contribution to the electron
self-energy that is absent in phonon self-energies. Moreover, a
sum rule analysis presented in Ref.
\raisebox{-1.2ex}{\Large\cite{Ole:sum_rule}} highlights in general
how electron and phonon self-energies may be qualitatively
different in strongly correlated systems.

\subsection{ARPES Experiments on Zn and Ni substituted Bi2212}
In this section, we comment on recent experiments about the
renormalization effects in Zn and Ni substituted Bi2212 crystal
\cite{Fink:Bi2212_Ni_Zn, TakaHashi:Bi2212_Ni_Zn}. The strength of
sharp renormalizations in these substituted crystals is found to
be weakened compared to the pristine crystals. Since the magnetic
properties are expectedly modified due to the Cu substitution by
these impurities, the authors concluded that the sharp
renormalization effects are induced by magnetic-related modes, not
phonons.

In fact, a close examination of the data published by V. B.
Zabolotnyy \emph{et al.} \cite{Fink:Bi2212_Ni_Zn} and K. Terashima
\emph{et al.} \cite{TakaHashi:Bi2212_Ni_Zn} implies that the
magnetic property is not the only modification due to the
substitution by Zn and Ni. First, although both sets of data are
consistent in the antinodal region where the strength of the band
renormalization is reduced, they are inconsistent with each other
on the kink strength along the nodal direction. In the data set of
V. B. Zabolotnyy \emph{et al.}, the kink strength is weaker in the
Zn or Ni doped samples, whereas there is no detectable change in
the data set reported by K. Terashima \emph{et al.}.

Second, the data from K. Terashima \emph{et al.} (Fig. 1(d)-(f) in
Ref. \raisebox{-1.2ex}{\Large\cite{TakaHashi:Bi2212_Ni_Zn}})
suggest that the bilayer-splitting structure is much clearer in
the pristine crystals than in the Zn and Ni doped crystals. Since
the authors have ruled out the possibility of a significant doping
level difference between pristine and impurity-doped crystals, the
distinct visibility variation of the bilayer structure implies a
impurity-related change in the electronic structure.

From these two observations on their data, it implies that not
only the magnetic properties could change, the band structure and
scattering behaviors could also be affected due to these
impurities. It is possible that these changes of the electronic
structures could ``weaken" the renormalizaton features observed in
the ARPES spectrum. Furthermore, we note that the strength of
electron-phonon coupling could also be modified by the substituted
impurities: this can be inferred from the change of the Fano
spectra lineshape of the B$_{1g}$ 340 cm$^{-1}$ phonon in Raman
spectral for Zn-doped YBCO\cite{Tajima:Zn_YBCO:Raman} and Th-doped
YBCO \cite{Hardy:Th_YBCO:Raman} resulting from an increase in the
phonon linewidth due to impurity scattering. Therefore, the
experiments on Ni and Zn substituted Bi2212 crystals are
inconclusive experiments to distinguish phonon and magnetic modes
as the origin of the renormalization effects.

\section{Conclusion}
We have shown that the temperature and doping dependence of the
renormalization effects, inter-band band scattering, and the
results of Zn and Ni doped materials can be understood in the
framework of electron-phonon coupling. On the other hand, the
issues that make it not plausible for the sharp kink being of spin
origin, especially the spin resonance mode, remain: i) the nearly
constant energy scale as a function of doping in small gap system
\cite{ZJZhou:multiple_mode}; ii) the multiple modes
\cite{ZJZhou:multiple_mode}; iii) the presence of clear kink in
the normal state
\cite{mode:sato:BiFamily,non:doping_dependence_monolayer,Lanzara:Noraml_state_PDH}
iv) the detailed agreement between B$_{1g}$ phonon based
explanation of the mode coupling as a function of momentum
\cite{Tanja:B1g:PRL,Tom:B1g:PRL}, while the spin resonance with
tiny spectral weight (2\%) is unlikely to give an explanation for
both nodal and antinodal renormalization; v) the accumulated
evidence for lattice polaron effect in underdoped and deeply
underdoped systems
\cite{Kyle:polaron:underdoped:phonon,Rosch:polaron:theory}. With
these weaknesses of the spin resonance interpretation, lattice
effect is a more plausible explanation of the renormalization
effects. It remains a possibility that the spin-fluctuation and
other strong correlation effects are also very important to
determine the electronic structure of cuprates; they likely
contribute to a smooth renormalization of the band and may be more
relevant to the higher binding energy. However, optical phonons
are the most probable origin for the renormalization effects due
to sharp modes near 40-70 meV, which is also supported by the
recent finding of STM experiments \cite{Davis:STM,Balasky:STM}.

\begin{acknowledgments}
W.S. Lee acknowledge the support from SSRL which is operated by
the DOE Office of Basic Energy Science, Division of Chemical
Science and Material Science under contract DE-AC02-76SF00515. T.
P. Devereaux would like to acknowledge support from NSERC, ONR
grant N00014-05-1-0127 and the A. von Humboldt Foundation.
\end{acknowledgments}

\appendix
\section{Migdal-Eliashberg based approach}

In the calculations presented herein, we evaluate electronic self
energies and spectral functions via Migdal-Eliashberg treatment,
as discussed in Ref. \raisebox{-1.2ex}{\Large\cite{Parks}}. The
dressed Green's function in the superconducting state is given in
Nambu notation by
\begin{widetext}
\begin{equation}\label{Eq:DressedG}
\hat{G}(\bk,\omega) = \frac{\omega Z(\bk,\omega)\hat{\tau}_0 +
[\epsilon(\bk)+\chi(\bk,\omega)]\hat{\tau}_3+\phi(\bk,\omega)\hat{\tau}_1}
{[\omega Z(\bk,\omega)]^2 - [\epsilon(\bk) + \chi(\bk,\omega)]^2 -
\phi^2(\bk,\omega)},
\end{equation}
from which the spectral
function follows $A({\bf k},\omega)=-\frac{1}{\pi} G_{1,1}^{\prime\prime}({\bf
  k},\omega)$ as shown in Figs. 1c,1d, and 2. The momentum-dependent components of the Nambu self
energy are given as generalizations of those found in Ref.
\raisebox{-1.2ex}{\Large\cite{Parks}}:
\begin{eqnarray}\label{Eq:wZ2} \nonumber
\omega Z_2(\bk,\omega) = \frac{\pi}{2N}\sum_{\bp,\nu}
|g_{\nu}(\bk,\bp-\bk)|^2 \big( [n_b(\Omega_{\nu}) +
n_f(E_\bp)][\delta(\omega + \Omega_{\nu} - E_\bp) +
\delta(\omega-\Omega_{\nu} + E_\bp)]\\ \quad\quad + [n_b(\Omega_{\nu}) +
n_f(-E_\bp)][\delta(\omega - \Omega_{\nu} - E_\bp) +
\delta(\omega+\Omega_{\nu}+E_\bp)] \big)
\end{eqnarray}
\begin{eqnarray}\label{Eq:chi2}\nonumber
\chi_2(\bk,\omega) =-\frac{\pi}{2N}\sum_{\bp,\nu} |g_{\nu}(\bk,\bp-\bk)|^2
\frac{\epsilon_\bp}{E_\bp}\big( [n_b(\Omega_{\nu}) +
n_f(E_\bp)][\delta(\omega + \Omega_{\nu} - E_\bp) -
\delta(\omega-\Omega_{\nu} + E_\bp)]\\ \quad\quad + [n_b(\Omega_{\nu}) +
n_f(-E_\bp)][\delta(\omega - \Omega_{\nu} - E_\bp) -
\delta(\omega+\Omega_{\nu}+E_\bp)] \big)
\end{eqnarray}
\begin{eqnarray}\label{Eq:phi2}\nonumber
\phi_2(\bk,\omega) = \frac{\pi}{2N}\sum_{\bp,\nu} |g_{\nu}(\bk,\bp-\bk)|^2
\frac{\Delta_\bp}{E_\bp} \big( [n_b(\Omega_{\nu}) +
n_f(E_\bp)][\delta(\omega + \Omega_{\nu} - E_\bp) -
\delta(\omega-\Omega_{\nu} + E_\bp)]\\ \quad\quad + [n_b(\Omega_{\nu}) +
n_f(-E_\bp)][\delta(\omega - \Omega_{\nu} - E_\bp) -
\delta(\omega+\Omega_{\nu}+E_\bp)] \big)
\end{eqnarray}
\end{widetext}
where $\nu$ denotes the phonon mode index, and $n_f$ and $n_b$ are
the Fermi and Bose occupation factors. $g_{\nu}(\bk,\bq)$ are the
corresponding electron-phonon couplings for mode $\nu$, given in
reference \cite{Tom:B1g:PRL} for the B$_{{1g}}$ and breathing
modes. We choose to model the $A_{1g}$ coupling via a momentum
independent coupling. Further details can be found in Ref.
\raisebox{-1.2ex}{\Large\cite{Screening:Tom}}.

\end{document}